\documentclass{article}[12pt]
\usepackage{amsmath,amssymb}
\usepackage{graphicx}%
\usepackage{verbatim}
\def\stackunder#1#2{\mathrel{\mathop{#2}\limits_{#1}}}

\newcommand{\Lee}[1]{\stackunder{#1}{\rm L}}

\newtheorem{assume}{Assumption}
\def\stackunder#1#2{\mathrel{\mathop{#2}\limits_{#1}}}
\newcommand{\pert}[2]{\stackrel{(#2)}{#1}}

\begin{document}

\begin{center}
{\bf \Large The Self-consistent Field Method and the Macroscopic\\[6pt] Universe Consisting of a Fluid and a Black Hole.\\[12pt]
Yu. G. Ignat'ev}\\
Physics Institute of Kazan Federal University,\\
Kremleovskaya str., 35, Kazan, 420008, Russia.
\end{center}

\begin{abstract}

The article discusses and substantiates a self-consistent approach to the macroscopic description of systems with gravitational interaction. Corrections to the equation of state of the fluid are found based on macroscopic Einstein equations which were obtained by averaging over microscopic spherically symmetric metric fluctuations created by the primary Black Holes in a fluid medium. It is shown that these corrections are effectively equivalent to  addition of a fluid to the system with the equation of state $p=-\varepsilon$. In addition, it is shown that, in this case, the mass of Black Holes can grow at a modern stage of evolution to very large values.\\

{\bf keywords}: Macroscopic cosmology, black holes, equation of state.

\end{abstract}

\section*{Introduction}
The foundations of the statistical theory of relativistic classical systems with gravitational interaction were stated in the early works of the Author \cite{YuI_GTO14}, \cite{YuI_GTO20}\footnote{see English version \cite{YuI_GC07}, details in monograph \cite{Yubook1}.}. A kinetic equation for massless particles in Friedmann's macroscopic world, accounting gravitational inte\-rac\-tion with microscopic spherically symmetrical local fluc\-tua\-tions of the metrics, which are generated by point sources of masses, was obtained on the basis of this theory in
\cite{YuI_Pop_IzvVuz}, \cite{YuI_Pop_ASS} . In these works, we took into account the cosmological evolution of spherical local fluctu\-ations generated by point masses that was examined for the case of ultra\-re\-la\-ti\-vis\-tic equation of matter in \cite{YuI_Pop_PhLet}, and for the case of arbitrary equation of state in \cite{Moroca1}, \cite{Moroca2}, \cite{Moroca3}.

Let us notice that the primary goal of the cited above works \cite{YuI_GTO14}, \cite{YuI_GTO20} was retrieval of the kinetic equation for photons in locally fluctuating Fried\-mann's world and associated problem of small-scale fluctuations of relict radiation. The Authors virtually did not pay attention to two important facts accompanied the procedure of this equation's derivation:
\vskip 8pt
\noindent \textbf{1}. the effective equation of state of gravitational fluctuations is not vaccum\footnote{it is incorrectly called an extremely rigid  in \cite{YuI_Pop_ASS}.}
\begin{equation}\label{dp+de=0}
\delta p+\delta\varepsilon =0;
\end{equation}
\noindent \textbf{2}.  the evolution law of point masses, which has direct relation to the problem of black holes.
\vskip 11pt

Taking into account cosmological and astro\-phy\-sical discoveries of recent year, in particular, the discovery of dark sector of cosmological matter, giant Black Holes possibly being located at many galaxies' centers, and other, we need to come back to research of the Universe filled with a fluid and singular massive sources. This paper is devoted to this problem.\footnote{This work was funded by the subsidy allocated to Kazan Federal University for the state assignment in the sphere of scientific activities.} We will consider classical model of the Universe filled with fluid with  \emph{barotropic} equation of state
\begin{equation}\label{barotrop}
p=k\varepsilon,
\end{equation}
where $k=\mathrm{Const}$ is a barotrope coefficient. We can't consider scalar field as a matter filling the Universe in this case. However we can account the cosmological constant. We hope to return to solving of the corresponding problem with scalar field in the nearest future. Due to importance of the considered problem we will strictly justify the suggested statistical approach and procedures of local fluctuations' averaging.

\section{The Self-Consistent Approach to Description of Metric's Local Fluctuations}
\vspace{-0.4cm}
\subsection{The Self-consistent Field Method}
\vspace{-0.2cm}

The statistical theory of derivation of {\it macroscopic Einstein equations} can be developed in a similar way as theory of many particles within the approach of {\it self-consistent field}, which initially appeared in celestial mechanics and later on was applied in theory of many particles (P. Weiss, 1907; I. Langmuir, 1913; I. Thomas, 1927; D. Hartree, 1928; V.A. Fock, 1930).
The special role in the development of the method of self consistent field belongs to A.A.Vlasov who in a series of his fundamental works  \cite{Vlasov1938} (1938) (see also \cite{Vlasov1945}, \cite{Vlasov}) has for the first time provided deep analysis of physical properties of charged particles of plasma, has shown inapplicability of gas-kinetic Boltzmann's equation to the description of plasma and proposed new kinetic equation of plasma (Vlasov equation) describing collective interaction of plasma's particles by means of the self-consistent field. Later on, the Vlasov's thepory was refined in  \cite{Land_Damp} (1946), strictly justified and generalized in works of N.N.Bogo\-loyubov  \cite{Bogolyubov46} (1946) and exquisitely applied by him to quantum statistics and theory of superfluidity \cite{Bogolyubov47} (1947). We also need to note the classical monograph by Chandrassekar \cite{Chandr} (1942), where on the basis of self-consistent field's method the principles of stellar dynamics were formulated and in essence, the theory of galactic structures was built.

According to the method of self-consistent field, one separate particle's motion can be described as a motion in a summary averaged field of the rest system's particles, neglecting influence of one particle's influence on the system's dynamics. The conditions of applicability of the method are long-range character of interparticle interactions and large number of interacting particles $N\gg1$. Simi\-larly we can consider the method with regard to pure nonlinear field systems. Here the separate small field mode that is characterized by certain field degrees of freedom, acts as a single particle and the condition of the method's applicability is a long-range character of the field and large number of its degrees of freedom. The gravitational interaction correspond to these conditions as well as possible; the conservation law of the full energy-mass and absence of negative  <<gravitational charges>> ensures its long-range character while large number of degrees of freedom is laid in the field nature of interaction itself. Therefore we can consider systems with gravi\-ta\-tional inte\-rac\-tion by the method of self - consistent field where each microscopic mode of gravitational per\-tur\-ba\-tion is small whereas the macroscopic self-consistent gravitational field is big.

\subsection{The Averaging of the Metric's Local Fluctuations}

Let us suggest that exact microscopic metrics of Riemann space - time $V_4$ can be written in form \cite{YuI_GC07,Yubook1}:
\begin{equation}\label{gik_micro}
g_{ik}(x)=\overline{g_{ik}(x)}+\delta g_{ik}(x),
\end{equation}
where $\overline{g}_{ik}(x)$ -- is a certain average macroscopic metrics corresponding to macroscopic space - time
$\bar{V}_4$, and $\delta g_{ik}(x)$ -- are microscopic fluctuations of the metrics so that it is
\begin{equation}\label{dg<<1}
\overline{\delta g_{ik}\delta g^{jk}}\ll 1,
\end{equation}
and
\begin{equation}\label{bar_dg=0}
\overline{\delta g_{ik}}=0.
\end{equation}
Let us use the overline to designate (here and further) a certain operation of averaging of the metrics and corresponding values, not defining it concretely yet. Let us just notice that this operation is quite a delicate procedure, significantly depending on the method of measurement (see details in \cite{YuI_GTO14,YuI_GTO20}). Let us also notice that <<pioneer>> methods of statistical averaging of the metrics being used by some researchers and, in essence, representing transition of methods of classical averaging by means of integrating metric values over spatial volume, are obviously inapplicable. First, as is known from Riemann's geometry, the result of integration of tensor over volume is an ambiguous operation and is not a tensor value. Second, since synchronization of observations is only possible in a synchronized system of reference, physical meaning of the result of integration of metric tenor over volume can't be defined as well. Third, macroscopic tool measuring the Universe's metrics, is simply non-realizable. Thus, following methods developed in \cite{YuI_GTO14,YuI_GTO20}, we average the metrics over certain arbitrary values, for instance, wave vectors, oscillations phases and suchlike. For example, in the papers cited above, the metrics was generated by massive particles and the averaging was carried out over coordinates of these particles not being arguments of tensor fields. Here we should impose certain requirements on {\it distribution function} $f(x^i,x^i_0)$, reflecting macroscopic features of symmetry of tensor fields being averaged.

Averaging the Einstein equations with the help of the described procedure, we find:
\begin{equation}\label{bar_einst_eq}
\overline{G^i_k(g)}=8\pi\overline{T^i_k(g,\phi_a)}+\Lambda \delta^i_k.
\end{equation}
As a result of nonlinearity of Einstein tensor and tensor of energy - momentum we have:
\begin{equation}\label{non_average}
\overline{G^i_k(g)}\not=G^i_k(\overline{g}); \quad \overline{T^i_k(g,\phi_a)}\not=T^i_k(\overline{g},\overline{\phi_a}),
\end{equation}
i.e., macroscopic source does not correspond to macroscopic metrics, as it turned to be the case in electrodynamics.

How should we obtain the equations that would describe the macros\-copic metrics in this case? The general principles of averaging of microscopic met\-rics in WKB - approximation and at derivation of macros\-copic Einstein equations were stated in the works of Isaacson \cite{Isaakson1,Isaakson2} (1968), and later used for the construction of the theory of relativistic statistical systems with gravitational interaction in the Author's works \cite{YuI_GTO14}, \cite{YuI_GTO20}, \cite{YuI_GC07}. So, let us perform averaging of the Einstein equations and corres\-pon\-ding equations of physical fields. Let us pay attention to the following circumstance. To begin the averaging we need to have a certain <<primeval>> start macroscopic metrics $g^{(0)}_{ik}(x)$ and certain pri\-meval start physical fields $\phi^{(0)}_a$, which from we can start the process of consecutive averaging iterations in a capacity of null approximation. These primeval fields satisfy a self-consistent system of field equa\-tions:
\begin{eqnarray}\label{Einst0}
G^i_k\bigl(g_{jm}^{(0)}\bigr)=8\pi T^i_k(\bigl(g_{jm}^{(0)},\phi^{(0)}_a\bigr)+\Lambda \delta^i_k;\\
\label{Field0}
\pert{\nabla}{0}_k T^k_i(g^{(0)},\phi^{(0)}_a)=0.
\end{eqnarray}
So, in accordance with common approach to averaging of gravitational fields let us consider a certain macroscopic Riemann space and let  $g^{(0)}_{ik}(x)\equiv \overline{g_{ik}(x)}$, $\phi^{(0)}_a(x) \equiv \overline{\phi_{a}(x)}$  be certain {\it yet unknown} macroscopic averages of the field values.

\begin{assume}\label{[d,int]=0}
Let us further assume that operations of differentiation or integration over coordinates are permutation with averaging operation:
\begin{eqnarray}\label{[diff,average]=0}
\overline{\partial_i\phi(x)}=\partial_i\overline{\phi(x)};\\
\label{[int,average]=0}
\overline{\int \phi(x)dx }=\int \overline{\phi(x)}dx  .
\end{eqnarray}
\end{assume}

Let us notice that the defined above operation of averaging fully satisfies the assumption \ref{[d,int]=0}.

Let it be:
\begin{equation}\label{dg,df}
\delta g_{ik}=g_{ik}-g^{(0)}_{ik};\quad \delta\phi_a=\phi_a-\phi^{(0)}_{a},
\end{equation}
-- are small local deviations of the metrics and physical fields from the averaged values, so that it is:
\begin{eqnarray}
\label{aver_g}
\overline{\delta g_{ik}}=0; & \overline{\partial_j\delta g_{ik}}=0;\\
\label{aver_phi}
\overline{\delta \phi}=0; & \overline{\partial_j \delta\phi}=0.
\end{eqnarray}

Let us expand the fields equations into Taylor series over smallness of metrics' and physical fields' deviations from the average values till second order over perturbations:
\begin{eqnarray}\label{Einst0-2}
G^{(0)}\ \!\!^i_k+G^{(1)}\ \!\!^i_k+G^{(2)}\ \!\!^i_k=\nonumber\\
8\pi\bigl( T^{(0)}\ \!\!^i_k+T^{(1)}\ \!^i_k+T^{(2)}\ \!\!^i_k\bigr)+\Lambda \delta^i_k,
\end{eqnarray}
and let us average now these equations with account of (\ref{bar_dg=0}), as well as linearity of operators $G^{(1)}_{ik}$ and $T^{(1)}_{ik}$
with respect to fluctuations:
\begin{eqnarray}\label{G1=0}
\overline{G^{(1)}\ \!\!^i_k(\delta g)}=G^{(1)}\ \!\!^i_k(\overline{\delta g})=0;\\
\label{T1=0}
\overline{T^{(1)}\ \!\!^i_k(\delta g,\delta\phi)}=T^{(1)}\ \!\!^i_k(\overline{\delta g},\overline{\delta\phi})=0.
\end{eqnarray}
Thus, in the first order of theory of perturbations we get the microscopic linear equations of the first order over smallness of metrics and physical fields
\begin{eqnarray}\label{evolut_dg}
G^{(1)}\ \!\!^i_k(\delta g)=8\pi T^{(1)}\ \!\!^i_k(\delta g,\delta\phi)+\Lambda \delta^i_k;\\
\label{evolut_df}
\nabla_k T^{(1)}\ \!\!^i_k(\delta g,\delta\phi)=0,
\end{eqnarray}
which we will call {\it the microscopic evolutionary equations for perturbations}.
Assuming\footnote{Let us notice that substitution (\ref{gik=gik0}) is a formal technique for metrics renormalization.} it is:
\begin{equation}\label{gik=gik0}
g^{(0)}_{ik}=\overline{g_{ik}},
\end{equation}
in the second order of perturbation theory after the averaging we obtain the macroscopic Einstein equations for macroscopic metrics
\begin{eqnarray}\label{Eq_<Einst2>}
G^{(0)}\ \!\!^i_k(\overline{g})=-\overline{G^{(2)}\ \!\!^i_k(\delta g)}+8\pi T^{(0)}\ \!\!^i_k(\overline{g},\overline{\phi})\nonumber\\
+8\pi\overline{T^{(2)}\ \!\!^i_k(\delta g,\delta\phi)} +\Lambda \delta^i_k,
\end{eqnarray}
according to which the macroscopic metrics in the second order as per the theory of perturbations is defined by Einstein equations with cosmological term and summary effective tensor of energy - momentum:
\begin{equation}\label{tilde_MET}
\tilde{T}^i_k=T^{(0)}\ \!\!^i_k(\overline{g},\overline{\phi})+\mathcal{T}^{(2)}\ \!\!^i_k,
\end{equation}
where it is
\begin{equation}\label{Tik_sum}
\mathcal{T}^{(2)}\ \!\!^i_k\equiv\overline{ T^{(2)}\ \!\!^i_k}-\frac{1}{8\pi}\overline{G^{(2)}\ \!\!^i_k}.
\end{equation}
Thus, the macroscopic Einstein equations of the second order over perturbations take the following standard form:
\begin{equation}\label{Eq_Einst_aver}
G^i_k(\overline{g})=8\pi \tilde{T}^i_k +\Lambda \delta^i_k.
\end{equation}
For the closure of macroscopic Einstein equations it is required to calculate macroscopic averages which are quadratic over local fluctuations of metrics and physical fields. These averages are defined by evolution equations (\ref{evolut_dg}) -- (\ref{evolut_df}).

\subsection{The Macroscopic Symmetries}

The following object is called a Lie derivative of $\Omega^i_k$ in the direction of vector field $\xi^i$  \cite{Petrov}.
\begin{equation}\label{Lee-det}
\Lee{\xi}U^i_k=\lim\limits_{dt\to 0}\frac{\Omega^i_k(x^i+\xi^i dt)-\Omega^i_k(x^i)}{dt},
\end{equation}
so that it is:
\begin{equation}\label{Lee_det}
\Lee{\xi}U^i_k=\xi^j\partial_j U^i_k-U^j_k\partial_j \xi^i+U^i_j\partial_k \xi^j.
\end{equation}
Particularly, if $U^i_k$ is a tensor and it is possible to replace partial derivatives in (\ref{Lee_det}) with covariant ones:
\begin{equation}\label{Lee_det_cov}
\Lee{\xi}U^i_k=\xi^j\nabla_j U^i_k-U^j_k\nabla_j \xi^i+U^i_j\nabla_k \xi^j
\end{equation}
this object is a tensor of the same valency as a source one.

Let us consider a pseudo-Riemann space $V_4$ as a primeval space which allows a certain group of motions $G^r$ with Killing vectors $\stackunder{(\alpha)}{\xi}$, defining a macroscopic symmetry:
\begin{eqnarray}\label{Lg0}
\Lee{\alpha}g^{(0)}_{ik}=0;\quad  \Lee{\alpha}\phi^{(0)}_{a}=0; \qquad   \bigl(\Lee{\alpha}\equiv \Lee{\xi_\alpha}\bigr).
\end{eqnarray}

Let us make the following assumption.
\begin{assume}\label{as2}
The macroscopic average of the Einstein tensor inherits the symmetry properties of the macroscopic metrics:
\begin{equation}\label{macr_sym}
\Lee{\xi}\overline{g_{ik}}=\sigma \overline{g_{ik}} \Longrightarrow \Lee{\xi} \overline{G_{ik}}= \sigma_1 \overline{g_{ik}},
\end{equation}
where $\Lee{\xi}$ -- is a Lie derivative (see, e.g. \cite{Petrov}), $\sigma(x),\sigma_1(x)$ -- are certain scalar functions.
\end{assume}

Let us notice that at $\sigma=0,\ \sigma_1=0$ we get a roup of motions $\bar{V}_4$, while at null values of these scalars we get a group of conformal transformations. Let us explain the assumption \ref{as2}. As is known from Riemann geometry (see \cite{Yubook1}), all geometrical objects inherit symmetry properties of metric tensor, for instance:
\begin{equation}\label{Lg}
\Lee{\xi}g_{ik}=0 \Rightarrow \Lee{\xi}\Gamma^j_{ik}=0;\; \Lee{\xi}R_{ijkl}=0;\; \Lee{\xi}T_{ik}=0.
\end{equation}
Hence it follows, as an example, that all symmetrical covariant tensor of the second valency have the same algebraic structure as a metric tensor. Thus it is logical to assume that algebraic structure of macroscopic tensors will also be the same.
Let us notice that since the following equalities are fulfilled:
\begin{equation}\label{LeeT0}
\Lee{\alpha}G^{(0)}\ \!^i_k=0; \quad \Lee{\alpha}\Lambda \delta^i_k=0; \quad\Lee{\alpha}T^{(0)}\ \!^i_k(\overline{g},\overline{\phi})=0,
\end{equation}
then, as a consequence of the assumption \ref{as2} the summary tensor of the energy-momentum of the second order over perturbations should also inherit the symmetries of the macroscopic metrics at averaging:
\begin{equation}\label{LeeT2=0}
\Lee{\alpha}\overline{\mathcal{T}^{(2)}\ \!^i_k}=0\Rightarrow \Lee{\alpha}\tilde{T}^i_k=0.
\end{equation}
Let us notice that assumption \ref{as2} imposes certain necessary conditions on  symmetry of scalar distribution function of random tensor fields $f(x^i,\vec{\lambda})$, in particular:
\begin{equation}\label{f_sym}
\Lee{\alpha}f(x^i,\vec{\lambda})=0.
\end{equation}
\subsection{The Macroscopic Einstein Equations of the Second Order}

Let us finally write out the system of equations defining macroscopic metrics in the second order of the perturbation theory. These equations consist of the system of linear equations for local perturbations of metrics and physical fields
\begin{eqnarray}\label{Einst1}
G^{(1)}\ \!\!^i_k(\delta g)=8\pi T^{(1)}\ \!\!^i_k(\delta g,\delta\phi);\\
\label{Field1}
\nabla_k T^{(1)}\ \!\!^i_k(\delta g,\delta\phi)=0
\end{eqnarray}
and macroscopic Einstein equations
\begin{eqnarray}\label{Eq2<Einst2>}
G\ \!\!^i_k(\overline{g})-\Lambda \delta^i_k=-\overline{G^{(2)}\ \!\!^i_k(\delta g)}+\nonumber\\
8\pi(T^{(0)}\ \!\!^i_k(\overline{g},\overline{\phi})+\overline{T^{(2)}\ \!\!^i_k(\delta g,\delta\phi)}),
\end{eqnarray}
where values $G^i_k(\overline{g})$ are calculated with respect to macroscopic metrics $\overline{g}_{ik}$.

Let us make the following important notice. Obtaining the macroscopic Einstein equations, we performed renormalization of metrics $g^{(0)}_{ik}\to
\overline{g}_{ik}$ . There are some consequences of that fact. First of all, according to the self-consistent field's method the macroscopic averages $\overline{G^{(2)}_{ik}(\delta g)}$ are not obligatory to be small as compared to $G_{ik}(\overline{g})$ since they are not a result of summation of infinite number of degrees of freedom of gravitational perturbations. Second, the cited renormalization of metrics leads to change $G_{ik}(g^{(0)})\to G_{ik}(\overline{g})$. This, in turn, means that macroscopic Einstein equations (\ref{Eq2<Einst2>}) are immediately resolved with respect to macroscopic metrics and not by the method of consecutive iterations:
$$\overline{g_{ik}}=g^{(0)}_{ik}+\overline{\delta g_{ik}}+\ldots$$
-- this is one of the main advantages of method of self-consistent field of Hartree - Fock - Vlasov - Bogolyuobov. The method of consecutive iterations would have pulled us in a complete different direction. To understand that, it is enough to imagine Einstein equations for the Friedmann Universe where the energy momentum tensor would have been defined by a gas of infinite number of  <<small photons>>. Following the method of consecutive approximations we would lay Minkovsky tensor in a capacity of null approximation. It is clear that we never would get a cosmological singularity with that approach, taking into account smallness of gravitational perturbations brought by <<small photons>> as compared to units of the Minkovsky metrics. This example pictorially shows that all the researchers involuntarily use the method of self-consistent field in the relativistic theory of gravitation, not making attempts to validate it.

%%%%%%%%%%%%%%%%%%%%%%%%%%%%%%%%%%%%%%%%%%%%%%%%%%

\section{The Self-Consistent Field Method for Macroscopic Friedmann Universe Consisting of a Fluid and Black Holes}
\subsection{Localized Spherically Symmetrical Perturbations in the Friedmann World}
Let us consider now small spherically symmetrical perturbations in a space - flat Friedmann Universe:
\begin{equation}\label{Freedman}
ds^2=a^2(\eta)[d\eta^2-dr^2 -r^2(d\theta^2+\sin^2\theta d\varphi^2)],
\end{equation}
assuming perturbed metrics in isotopical coordinates in the following form (see, e.g., \cite{Land_Field}):
\begin{equation}\label{pert_fredman}
ds^2=\mathrm{e}^\nu d\nu^2-\mathrm{e}^\lambda[dr^2 -r^2(d\theta^2+\sin^2\theta d\varphi^2)],
\end{equation}
where $\nu(r,\eta),\lambda(r,\eta)$. The world line of a massive particle is a line of time where mass of particle $m(\eta)$ is yet an arbitrary function (details see in \cite{YuI_Pop_ASS}). Assuming $m, v, \nu',$ $\lambda'$ are values of first order of smallness, ($v(r,\eta)$ is a radial velocity of fluid's motion, $\phi'\equiv \partial\phi/\partial r$). In the linear approximation one of the Einstein equation we get the following consequence\footnote{In \cite{Yu19}, a typo was made in signs before $\xi$ in \eqref{lambda+nu}, as a result of which an incorrect formula for the Schwarzschild metric was obtained. However, this error does not affect further results.}
\begin{equation}\label{lambda+nu}
\lambda=\ln a^2-\xi(r,\eta);\quad \nu=\ln a^2 +\xi(r,\eta), \quad \xi\ll1.
\end{equation}
The linearized system of Einstein equations takes form ($\dot{\phi}\equiv \partial\phi/\partial \eta$):
\begin{eqnarray}
\label{P,E}
3\frac{\dot{a} ^2}{a^4}=8\pi\varepsilon_0;\quad 2\frac{\ddot{a}}{a^3}-\frac{\dot{a} ^2}{a^4}=-8\pi p_0;\\
\label{ddot_xi}
\ddot{\xi}+3\frac{\dot{a}}{a}\dot{\xi}+\left(2\frac{\ddot{a}}{a}-\frac{\dot{a} ^2}{a^2}\right)\xi=-8\pi a^2 \frac{dp_0}{d\varepsilon_0}\delta\varepsilon;\\
\label{xi''}
-\frac{1}{r^2}\frac{\partial}{\partial r}\left(r^2\frac{\partial \xi}{\partial r}\right)+3\frac{\dot{a}}{a^2}\frac{\partial}{\partial \eta}(a\xi)
=8\pi\delta\varepsilon +8\pi\frac{m}{a^3}\delta(\mathbf{r});\\
\label{dv/d_eta}
v=\frac{1}{8\pi(\varepsilon_0+p_0)}\frac{\partial}{\partial \eta}(a\xi').
\end{eqnarray}
Equations \eqref{P,E} describe non-perturbed models of Friedmann Universe, equation \eqref{dv/d_eta} defines radial velocity of a fluid in localized perturbation. Two remaining equations \eqref{ddot_xi} and \eqref{xi''}, one of which is singular, define two functions of two variables, $\xi(r,\eta)$, $\delta\varepsilon$ and one function of time $m(\eta)$. To single out singular part of the solution, we put:
\begin{equation}\label{xi,m,psi}
\xi=\frac{2}{r}(m-\psi(r,\eta)),
\end{equation}
where $m=m(\eta)$ and
\begin{equation}\label{lim_psi}
\lim\limits_{r\to0}\frac{\psi(r,\eta)}{r}<\infty.
\end{equation}
The local perturbation's metric here \eqref{pert_fredman} in accordance with \eqref{lambda+nu} takes the following form:
\begin{eqnarray}\label{lin_Shvarc}
ds^2\approx a^2(\eta)\biggl(d\eta^2(1-2(m-\psi)/r) -\frac{dl^2_0}{1-2(m-\psi)/r}\biggr),
\end{eqnarray}
where $dl^2_0=dr^2 +r^2(d\theta^2+\sin^2\theta d\varphi^2)$  -- is a 3-dimensional Euclidian metrics of a space, con\-for\-mally corresponding to the Friedmann space.\\
Singling out singular part in the equation \eqref{xi''} and separating variables in non-singular parts of equations \eqref{ddot_xi} and \eqref{xi''}, let us obtain autonomous equations on functions $\xi(r,\eta)$ and $m(\eta)$;
\begin{eqnarray}
\label{ddot_m}
\ddot{m}+\frac{\dot{a}}{a}\dot{m}\left(1+3\frac{dp_0}{d\varepsilon_0}\right)+m\left(\frac{\ddot{a}}{a}-2\frac{\dot{a} ^2}{a^2}\right)=0;\\
\label{eq_psi}
\ddot{\psi}-\psi''\frac{dp_0}{d\varepsilon_0}+\frac{\dot{a}}{a}\dot{\psi}\left(1+3\frac{dp_0}{d\varepsilon_0}\right)+\nonumber\\
\psi\left(\frac{\ddot{a}}{a}-2\frac{\dot{a} ^2}{a^2}\right)=0.
\end{eqnarray}
The equation \eqref{eq_psi} with respect to nonsingular func\-tion $\psi(r,\eta)$ at $dp_0/d\varepsilon_0 >0$ is a hyperbolic one.

If we like to consider namely \emph{localized} per\-tur\-ba\-tions, which appeared as a result of radial redistribution of the Friedmann matter rather than external masses addition, taking into account the fact that per\-tur\-ba\-tions of density spread with the speed of sound $v_s=\sqrt{dp_0/d\varepsilon_0}$ we have to require:
\begin{equation}\label{xi_8}
\left.\xi(r,\eta)\right|_{r=r_s(\eta)}=0;\quad \left.\xi'(r,\eta)\right|_{r=r_s(\eta)}=0,
\end{equation}
where $r_s(\eta)$ is a sound horizon.
Regarding the function $ \ psi $, these boundary conditions take the form:
\begin{equation}\label{psi_8}
\left.\psi(r,\eta)\right|_{r=r_s(\eta)}=m(\eta);\quad \left.\psi'(r,\eta)\right|_{r=r_s(\eta)}=0.
\end{equation}
Thus, outside the hypersurface of a sound horizon $r\geqslant r_s(\eta)$ local fluctuations completely disappear and the Universe is ``not aware'' of their existence\footnote{see below}. In the proximity of singularity $r\to0$ according to \eqref{lim_psi} it is $\psi\to 0$, and metrics \eqref{lin_Shvarc} takes the form of Schwarzschild metric with a mass of a central point particle $m(\eta)$. This mass appears due to radial redistribution of matter inside the sphere of sound horizon as a result of 2 competing processes: accretion and evaporation of matter. In this sense, the local spherical perturbation is a \emph{linear} model of a black hole.

Let us consider solutions of equations \eqref{ddot_m} and \eqref{eq_psi} in two cases being of interest to us\footnote{Solutions for other equations of state, among them - automodeling solutions see in \cite{Moroca1} -- \cite{Moroca3}}.
\subsubsection{Ultrarelativistic Equation of State $p_0=\varepsilon_0/3$}
Taking into consideration the known solution $a\sim \eta$ on this stage of expansion let us write out equations \eqref{ddot_m} and \eqref{eq_psi} for this case:
\begin{eqnarray}\label{ddot_m1}
\ddot{m} +\frac{2}{\eta}\dot{m}-\frac{2}{\eta^2}m=0;\\
\label{eq_psi1}
\ddot{\psi}-\frac{1}{3}\psi''+\frac{2}{\eta}\dot{\psi}-\frac{2}{\eta^2}\psi=0.
\end{eqnarray}
The equation \eqref{ddot_m1} is easily integrated, its solution contains growing and decreasing modes which was mentioned above:
\begin{equation}\label{m(eta)}
m(\eta)=\mu_0\frac{\eta}{\eta_0}+\mu_1\frac{\eta^2_0}{\eta^2},
\end{equation}
where $\mu_0,\mu_2$ are certain constants so that $m_0=m(\eta_0)=\mu_1+\mu_2$.

To resolve the equation \eqref{eq_psi1} let us make a change $\eta=\sqrt{3}\tau$ and
\[\psi=\frac{\partial}{\partial \tau}\frac{\Phi(r,\tau)}{\tau}. \]
Then we find for a solution of the equation \eqref{eq_psi1} (details see in \cite{Yubook1})
\[\Phi=\Phi_+(r+\tau)+\Phi_-(r-\tau),\]
where $\Phi_\pm$ -- are certain arbitrary function satisfying stated above boundary conditions.

Concrete solutions also contain both growing and decreasing modes corresponding to known\\ Lifshitz solutions for scalar perturbations of the Friedmann Universe (see \cite{Lifshitz,Land_Field}). We will not bring here the cumbersome formulas and limit ourselves to consideration of a particular case -- the account of the growing mode of perturbations. We also preserve only growing mode in formula for mass of a singular source \eqref{m(eta)}, assuming further $\mu_1=0$. Thus, $\mu_0$ is a mass of a singular source in the time instant $\eta=\eta_0$ of switching of the ultrarelativistic stage of expansion:\footnote{Let us notice that in modern version of the standard cosmological model the ultrare\-la\-ti\-vistic stage follows after the inflationary one. In this case \eqref{m0(eta)} can contain also a decreasing mode $\sim\eta^{-2}$. We will return to this question in the nearest future.}. So, let us find for the ultra\-re\-la\-ti\-vistic stage of the Universe expansion:
\cite{Yubook1}
\begin{eqnarray}\label{m0(eta)}
m(\eta)=\mu_0 \frac{\tau}{\tau_0};\\
\label{xi0(eta)}
\xi(r,\eta)=\frac{2\mu_0}{r}\left[1-\frac{r}{2\tau_0}\left(3-\frac{r^2}{\tau^2}\right)\right]\mathrm{U}_+(\tau-r),
\end{eqnarray}
where $\tau=\eta/\sqrt{3}$ -- is a time variable in a sound scale, $\mathrm{U}_+(z)$ -- is a Heavyside function.
\subsubsection{Non-Relativistic Equation of State $p_0=0$}
In this case the equations of state \eqref{ddot_m} and \eqref{eq_psi} are easily integrated:
\begin{eqnarray}\label{m1}
m(\eta)=\mu_1\left(\frac{\eta_1}{\eta}\right)^3+\mu_2\left(\frac{\eta}{\eta_1}\right)^2; \\
\label{xi1}
\quad \psi=\psi_1(r)\left(\frac{\eta_1}{\eta}\right)^3+\psi_2(r)\left(\frac{\eta}{\eta_1}\right)^2,
\end{eqnarray}
where $\eta_1$ -- is an instant of change of the equation of state, $\psi_1(r)$ and $\psi_2(r)$ -- are arbitrary functions. Sewing together the increasing mode of this solution and solution
\eqref{m0(eta)} -- \eqref{xi0(eta)}, we find:
\begin{eqnarray}\label{m1(eta)}
m(\eta)=& \displaystyle \mu_0 \frac{\eta^2}{\eta_0\eta_1};\\
\label{xi1(eta)}
\xi(r,\eta)=& \displaystyle \frac{2\mu_0\eta^2}{r\eta_0\eta_1}\left[1-\frac{3\sqrt{3}r}{2\eta_1}\left(1-\frac{r^2}{\eta_1^2}\right)\right]\\
& \displaystyle \times\mathrm{U}_+(\tau_1-r),\nonumber
\end{eqnarray}
\subsection{The Averaging of Local Fluctuations}
Let now in the Friedmann world there is not one but many massive identical particles with coordinates $\mathbf{r}_a=\{x_sa,y_a,z_a\}$. In the approximation linear over $m$, the summary metrics of the space - time can be written in a form of superposition of local fields:
\begin{equation}\label{g+h}
ds^2=(g_{ik}+a^2 h_{ik})dx^idx^k
\end{equation}
where $g_{ik}$ -- is a Friedmann metrics and
\begin{equation}
h_{ik}=-a^2\delta_{ik}\sum\limits_{a} \xi_a(|\mathbf{r}-\mathbf{r}_a|).
\end{equation}
Let now the coordinates of massive particles take arbitrary values and correlation between positions of separate particles is absent  and let
$N$  be an average number of massive particles which are in conformal volume $V=4/3\pi r_0^3$, i.e. for volume of Minkovsky metrics, conformally corresponding to Friedmann metrics \eqref{Freedman}, where $r_0=\tau_0\equiv \eta_0/\sqrt{3}$ -- is a sound horizon on the time instant of change of the ultrarelatvistic phase of expansion to non\-re\-la\-ti\-vistic. For simplicity let us assume masses of these particles $m(\eta)$ are the same, i.e. the particles are identical. In the opposite case they should have been subject to averaging over particles' mass spectrum, i.e. over parameter $\mu_0$. Let us calculate the averages of the values by the rule:
\begin{equation}\label{average}
\overline{\phi(x)}=\prod\limits_a \frac{1}{V_a} \int d^3 \mathbf{r}_a \phi(\mathbf{r}|\mathbf{x}_1,\ldots,\mathbf{x}_N).
\end{equation}
Let us find, calculating the average values in such way:
\begin{equation}
\label{<h>}
\overline{h_{ik}}=-a^2\delta_{ik} \frac{3\mu N}{5r_0}=a^2\cdot \mathrm{Const}.
\end{equation}
Carrying out renormalization of metrics in accor\-dance with the above cited examples, we find for the renormalized potential $\xi_a\to \xi_a-3\mu/5r_0$, where averages from renormalized fluctuations now equal to zero and renormalization of the macroscopic metrics is reduced to multiplying it to constants, i.e., is removed by scale transformation. Using the renormalized value of the function $\xi(r)$ we can assure that change of the full mass inside raidus $r_0$ is strictly equal to zero.

Next, let us find the averages:
\begin{eqnarray}\label{<xi^2>}
\overline{\xi^2(\mathbf{r})}=\frac{108}{175}N\left(\frac{2\mu_0}{r_0}\right)^2;\\
\label{h_{ik}h_{lm}}
\overline{h_{ik}h_{lm}}=\delta_{ik}\delta_{lm}a^4\overline{\xi^2};\\
\label{d_jhH_ik}
\overline{\partial_j h_{ik}\cdot h_{lm}}=0;\quad \overline{\partial_4 h_{ik}\partial_j h_{lm}}=0.
\end{eqnarray}
The averages of form $\overline{\partial_\alpha h_{ik}\partial_\beta h_{lm}}$ diverge as all potential values proportionally to $1/r$; the diver\-gence of these values is associated with known divergence of the self-energy. Performing standard renor\-ma\-li\-zation of mass, we obtain:
\begin{equation}\label{d_ad_b}
\overline{\partial_\alpha \xi \partial_\beta\xi} =\delta_{\alpha\beta} \frac{6\pi N}{r^2_0}\left(\frac{2\mu_0}{r^2_0}\right).
\end{equation}

\subsection{The Effective Tensor of Energy - Momentum}
Calculating now the averages from corrections to Einstein tensor, caused by local fluctuations of metrics, let us find the correction to tensor of energy - momentum of Friedmann's dust:
\begin{equation}\label{dT}
\delta T^i_k=-\frac{1}{8\pi}\delta \overline{G^{(2)}\!\ ^i_k}=\frac{9N}{4r^2_0}\left(\frac{2\mu_0}{r_0}\right)^2 \delta^i_k.
\end{equation}

\section{Conclusion}
Comparing the obtained result with the expression for the energy - momentum tensor of the ideal fluid
\begin{equation}\label{EMT}
T^i_k=(\varepsilon+p)u^iu_k - p\delta^i_k,
\end{equation}
we can draw the following conclusions.
\vskip 11pt
\noindent \textbf{1}. The correction to tensor of the energy - momentum with respect to locally spherically symmetrical fluc\-tua\-tions of metrics has a form of energy - momentum tensor of the ideal fluid with vacuum equation of state:
\begin{equation}\label{p+e=0}
\varepsilon_g + p_g=0.
\end{equation}
\vskip 11pt
\noindent \textbf{2}. The pressure of this fluid is negative and constant:
\begin{equation}\label{p_g<0}
p_g=-\frac{9N}{4r^2_0}\left(\frac{2\mu_0}{r_0}\right)^2.
\end{equation}
\vskip 11pt
\noindent \textbf{3}. Thus, the correction to tensor of the energy - momentum caused by root-mean-square fluctuations of metrics can be interpreted as cosmological constant or as a component of dark matter in the classical ideal fluid. As a consequence of constancy of this term, it should lead to inflationary expansion of the Universe at late stages of the evolution.

Let us also notice that formula of growth of local \emph{point} masses  \eqref{m1(eta)} can be interpreted as a formula of growth of the Black Holes as a result of cosmological evolution. On the non-relativistic stage of the extension this gives us the growth law of the mass $m\sim t$.

To understand the physical meaning of the found result it is required to consider more in details the described above and in the Author's previous papers \cite{YuI_GTO14} -- \cite{YuI_Pop_ASS} method of self-consistent field with respect to systems with gravitational interaction. As a consequence of the Birkhoff theorem we cannot bring an external mass to the Friedmann world otherwise this world would become heterogenous. Consequently, the microscopic perturbations should not change the macroscopic energy density of the Friedmann Universe. The considered above local perturbations exactly do satisfy this principle. Hence it follows that the basic Einstein equation for the macroscopic Universe cannot change with account of perturbations not to mention the change of a global parameter such as curvature. In a nutshell, mass of particles is already accounted in the mac\-ros\-copic equations of the Universe's evolution -- it does not change  its macroscopic energy density. Renormalization of metrics, being quite a delicate procedure, should be carried out with an account of this fact. The single thing which could be influenced by localized fluctuatinos in an isotropic Universe is macroscopic equation of state. Here our theory inherits all main characteristics of the method of self-consistent field for electrodynamical systems. The difference appears only in one aspect: a closed statistical system should be electroneutral as a whole. This leads to the well-known Debye shielding of a charge in systems of charged particles. In the case of gravitation all particles' masses have like ``charges'' which originate macroscopic Friedmann background. This difference of gravitational systems is actually being managed by Birkhoff theorem which limits the action of local gravitational per\-tur\-ba\-tions. The results of this paper should be understood with the account of this fact. The fact that quadratic correction to the pressure should be negative i.e., the pressure in the system with interparticle gravitational attraction should be less than in a homogenous liquid with the same energy density, is physically obvious. The strict equality  \eqref{p+e=0} is less obvious.

The Author expresses his gratitude to prof. Vitaly Melnikov for the in-time support of the research in this direction.
%%%%%%%%%%%%%%%%%%%%%%%%%%%%%%%%%%%%%%%%%%%%%%%%%%%%%%%%%%%%%%%%%

%

\end{document}